\documentclass[10pt]{iopart}

\usepackage{float} % For [H] placement
\usepackage{placeins} % For \FloatBarrier
\usepackage{lineno}
\usepackage{subcaption}
\usepackage{cite}
\usepackage{amsmath,amssymb,amsfonts}
\usepackage{graphicx}
\begin{document}

\title[Anyonic Josephson junctions: Dynamical and
ground-state properties]{Anyonic Josephson junctions: Dynamical and
ground-state properties}

% \author{Jessica John Britto \dag}

% \address{\dag\ Raman Research Institute, C. V. Raman Avenue, Sadashivanagar, Bangalore 560080, India}
% \address{\dag\ Department of Physics, Indian Institute of Technology Kharagpur, Kharagpur 721302, India}
\author{Jessica John Britto}

\address{Raman Research Institute, C. V. Raman Avenue, Sadashivanagar, Bangalore 560080, India}
\address{Department of Physics, Indian Institute of Technology Kharagpur, Kharagpur 721302, India}

% \ead{jessicajohnbritto@kgpian.iitkgp.ac.in}
\vspace{10pt}
\begin{indented}
\item[{June 21, 2025}]
\end{indented}

\begin{abstract}
Bosons with density-dependent hopping on a one dimensional lattice have been shown to emulate anyonic particles
with fractional exchange statistics. Leveraging this, we construct
a Josephson junction setup, where an insulating barrier in the
form of a Mott-insulator is sandwiched between two
superfluid phases. This is obtained by spatially varying either
the statistical phase or the strength of the on-site interaction
potential on which the ground state of the system depends.
Utilizing numerical methods such as exact diagonalization and
density renormalization group theory, the ground state properties
of this setup are investigated to understand the Josephson effect
in a strongly correlated regime. The dynamical properties of this
model for different configurations of this model are analyzed to
find the configurations that can produce the Josephson effect.
Furthermore, it is observed that continuous particle flow
over time is achievable in this proposed model solely by creating
an initial phase difference without any external biasing.
\end{abstract}

\vspace{2pc}
\maketitle
\ioptwocol

\section{Introduction}
\label{sec intro}
Recent experiments have realized anyons in 1D optical lattices using conditional-hopping bosons via assisted Raman tunneling \cite{Keilmann:2010cm, greschner2015anyon}, with demonstrations including anyonic random walks \cite{kwan2024realization}. Motivated by these studies, we study the Josephson effect in a strongly correlated many-body regime. We propose a multi-site anyonic Josephson junction model, designed based on the ground-state properties of the 1D anyonic Hubbard model, to characterize this phenomenon. We analyze ground-state and dynamical properties of the multi-site anyonic Josephson junction using observable quantities.

First, we briefly describe basic differences among anyons, bosons and fermions, followed by 1D anyonic Hubbard model in Section \ref{subsec anyonic hubbard model}. In Section \ref{sec anyonic josephson junction}, the ground-state properties of this model for a system size of 64 lattice sites, are analyzed. 
% to find valid configurations that can potentially produce the Josephson effect under Hamiltonian time evolution. 
In Section \ref{sec dynamics of AJJ}, the dynamics of the model under different configurations are studied starting from the two-site model \cite{brollo2022two}, up to six lattice sites (and 64 lattice sites), to identify configurations showing Josephson effect observed in conventional Josephson junctions. Furthermore, the analysis of the dynamical properties shows that by creating phase difference between two parts of the given system, results in continuous current flow without any external biasing. 
% \textcolor{red}{In the main portion of the article, only observables aligning strongly with the predictions are considered. Nevertheless, in the appendices of this article, there are more observables that I have used to verify the results further.}  
\begin{figure}[ht]
    \centering
    \includegraphics[width=0.8\linewidth]{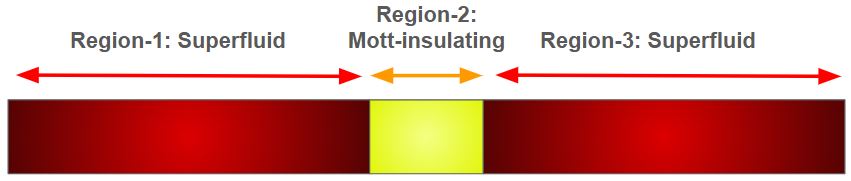}
    \caption{Anyonic Josephson junction general setup}
    \label{fig:Anyonic Josephson junction general setup}
\end{figure}

\subsection{Anyons, Bosons and Fermions}
\label{subsec anyons, bosons and fermions}
In three dimensions or higher, quantum particles only follow Fermi-Dirac and Bose-Einstein statistics \cite{lerda2008anyons}. Anyons are quasi-particles that are found only in $<=2$ dimensions \cite{lerda2008anyons}, although their properties can be modeled in arbitrary dimensions as formulated by Haldane \cite{brollo2022two, Keilmann:2010cm}. Anyons interpolate between the particle statistics of bosons and fermions. Table \ref{table - anyons, bosons and fermions} summarizes some of the key differences of bosons, fermions and anyons. 
\begin{table}
    \centering
    % \begin{tabular}{|p{3cm}|p{3cm}|p{3cm}|p{3cm}|}  \hline
    \begin{tabular}{|p{2.4cm}|p{1.8cm}|p{1.95cm}|p{1.8cm}|}  \hline 
        \textbf{Property} & \textbf{Bosons} & \textbf{Fermions} & \textbf{Anyons} \\ \hline 
        \small Spin & \small Integer (e.g., 0, 1, 2) & \small Half-integer (e.g., 1/2, 3/2) & \small Fractional \\ \hline 
    \small Wavefunction under exchange & \small Symmetric & \small Anti-symmetric & \small $e^{i\theta}$ \\ \hline 
    \small Exchange Phase $\theta$ & \small $\theta = 0$ & \small $\theta = \pi$ & \small $0 < \theta < \pi$ \\ \hline
    \end{tabular}
    \caption{Comparison of bosons, fermions \& anyons}
    \label{table - anyons, bosons and fermions}
\end{table}

As given in \cite{Keilmann:2010cm}, anyonic operators can be mapped to bosonic operators and the mapping is given in \ref{eq anyons-bosons mapping} where $b_{i}^{\dagger} (a_{i}^{\dagger})$, $b_{i} (a_{i})$ are the bosonic (anyonic) creation and annihilation operators, $n_{i} = b_{i}^{\dagger}b_{i} = a_{i}^{\dagger}a_{i}$, and $\theta$ is the statistical phase.  
\begin{equation}
	\label{eq anyons-bosons mapping}
	a_{j} = b_{j} \, \exp\left( i \theta \sum_{i=1}^{j-1} n_{i} \right)
\end{equation}
This mapping is non-local as it depends on a string of $n$ operators of other lattice sites to construct an anyonic annihilation operator for a lattice site $j$. 
\begin{equation}
	\label{anyon_commutation}
	\begin{split}
	a_{j}a^{\dagger}_{k} - e^{-i\theta sgn(j-k)}a_{k}^{\dagger}a_{j} & = \delta_{jk} \\
	a_{j}a_{k} & = e^{i\theta sgn(j-k)}a_{k}a_{j}
	\end{split}
\end{equation}
From the commutation rules for anyons, anyons with $\theta = \pi$, we obtain pseudofermions which means two such particles act as fermions off-site and as bosons on-site. This is because for the on-site case, we have $(j-k) = 0$ which results in the bosonic commutation rules irrespective of the value of $\theta$.

\subsection{Anyonic Hubbard model}
\label{subsec anyonic hubbard model}
The 1D anyonic Hubbard model has been studied previously in \cite{Keilmann:2010cm, greschner2015anyon}. In this section, key findings of the 1D anyonic Hubbard model are summarized, when it is expressed in terms of bosonic operators using the mapping given in \ref{eq anyons-bosons mapping}. We study this model specifically for unit filling case (i.e, $N_{sites} = N_{particles}$) with fixed total number of particles. 
\begin{equation}
\label{AHMHamil}
	H_{AHM} = -J\sum_{i} a_{i}^{\dagger}a_{i+1} + \frac{U}{2} \sum_{i}n_{i}\cdot(n_{i}-1)
\end{equation}
    \begin{equation}
	\label{boson_HamilAHM}
	H_{AHM}^{b} = - J \sum_{j}^{N_{sites}} \left(b_{j}^{\dagger} b_{j+1} e^{i\theta n_{j}} + \text{h.c}\right) + \frac{U}{2} \sum_{j}^{N_{sites}} n_{j} (n_{j} - 1)
\end{equation}

\begin{figure}
    \centering
    \includegraphics[width=0.9\linewidth]{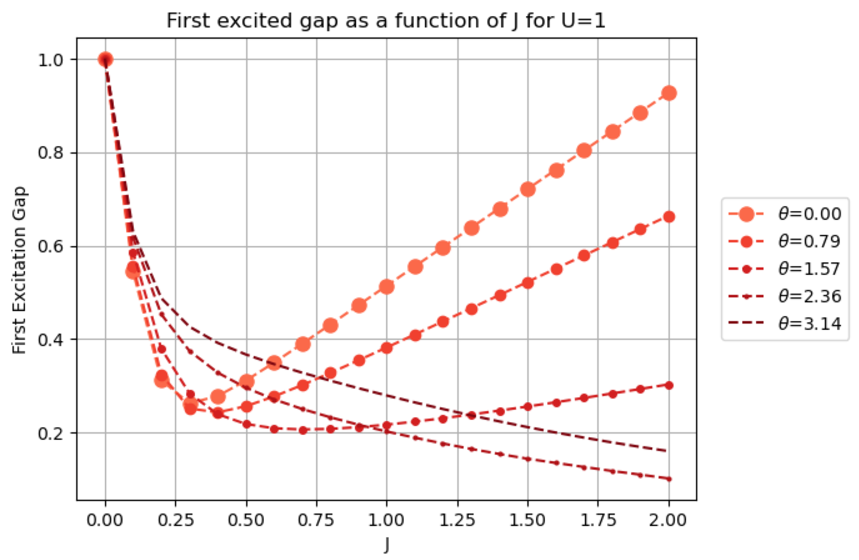}
    \caption{First excitation gap as a function of hopping amplitude $J$ for different values of statistical phase $\theta$}
    \label{fig:first excitation gap AHM}
\end{figure}

For a constant value of $\frac{U}{J}$, with increase in $\theta$ from zero to $\pi$, the system tends to show more prominent properties of the Mott-insulating phase. Therefore, for $\theta = \pi$, the model seems to act as a Mott-insulator as given in \cite{Keilmann:2010cm}. The first-excitation gap for this model validates this statement as shown in \ref{fig:first excitation gap AHM} since with increase in $\theta$ from zero to $\pi$, the critical value of $J$ ($J_{crit}$) also increases. Therefore, $J_{crit}$ for the statistical phase value $\pi$, it is expected to take an infinite value based on the plot given in \ref{fig:first excitation gap AHM}.

For a constant $U = 1$, the critical value of the hopping amplitude for the superfluid-Mott-insulating transition $J_{crit}$ increases with increase in $\theta$. The initial decrease in the plot \ref{fig:first excitation gap AHM}, before crossing $J_{crit}$ can be understood by doing Taylor's series expansion on the excitation gap in terms of small $J$ values compared to $U$. For a unit-filling case in the Mott-insulating phase, the ground-state consists of one particle per site. Therefore, this gap should be equal to $U=1$ as it is the minimum energy required to move a particle from one site and put it with another to create the minimum excitation. This reasoning holds for all values of $\theta$ for $J=0$. 

Beyond $J_{crit}$, the ground-state of this system tends to show superfluid properties where the particles should experience no on-site interactions ideally. In this case, we can think of this model in the tight-binding limit except this model will have an additional phase for $\theta \neq 0$ as given in \ref{boson_HamilAHM}. Since this is a finite system of size 64, the excitation gap is finite, however, the gap will vanish for large system size $\rightarrow \infty$.
%% Omit the explanation for density profile %%
% This can be understood from the bulk sites of the model through its density profile for a constant $U$ and $\theta = \pi$ (pseudofermions), where on increasing the hopping amplitude, $J$, the average number of particles on each site is exactly one while for $\theta = 0$ (bosons), is more than one, indicating the system's preference to be in superfluid phase. Note that in this model, the anyons are composite system of hopping density-dependent bosons, therefore, their occupational rule of anyons generated in this way are same as those of bosons or fermions. This can be understood from its first excitation gap where with increase in $J$.

Note that even in the absence of on-site interaction potential, the phase term $e^{i\theta n_{j}}$ associated with the hopping term in eq. \ref{boson_HamilAHM} induces complex many-body interactions. These interactions driven by the statistical phase $e^{i\theta n_{j}}$, provide a key motivation for studying the anyonic Josephson junction as a platform for exotic quantum dynamics.

\section{Anyonic Josephson junctions}
\label{sec anyonic josephson junction}
In this section, the ground-state properties of a multi-site anyonic Josephson junction are studied for a unit-filling model where $N_{sites} = N_{particles}$. This model is studied under different configurations which is achieved by first setting $\theta = 0$ and varying $U$ across the three regions to create superfluid and Mott-insulating phases as shown in \ref{fig:Anyonic Josephson junction general setup}. Secondly, the same can be obtained by setting $U$ to a constant while varying $\theta \in \{0, \pi\}$. The main reason behind this is to create a very strong superfluid or Mott-insulating phase so that this creates a simple scenario to study the Josephson effect in a strongly-correlated regime\footnote{Additionally, in these cases, we considered only $\theta = \{0,  \pi\}$ as during our numerical simulations, we noticed that MPS based DMRG algorithms were more efficient in simulating time-evolution of the given model as their Hamiltonian was real. Simulation of time-evolution was time-consuming for values of $\theta$ in between $0$ and $\pi$. Therefore, we could not run the simulations several times for different cases to verify the correctness of the results for $\theta \in (0, \pi)$. Block2 software package \cite{zhai2023block2} was used for the simulations in section \ref{sec anyonic josephson junction}.}.  Each configuration analyzed in this section has $64$ lattice sites: 30, 4, 30 sites in regions-1, 2 and 3 respectively, unless mentioned otherwise.

\subsection{Type-1: \protect$U_{1} = U_{3} < U_{2}$}
\label{subsec type1 BJJ}
%% Add configuration image %%
In this type, $\theta=0$ in all the three regions while $U$ is varied with $U_1=U_3$ thereby making this setup symmetry about region-2. This configuration can be considered a bosonic Josephson junction as only bosons are being used.  
\begin{figure}[ht]
    \centering
    \includegraphics[width=0.8\linewidth]{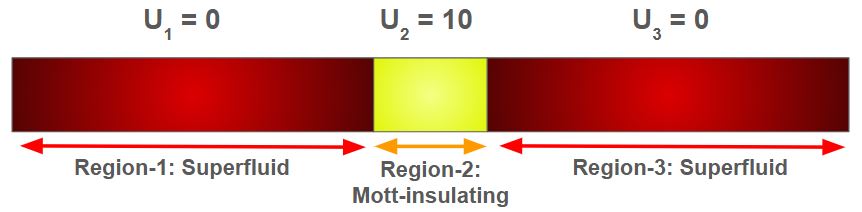}
    \caption{\protect$U_{1} = U_{3} < U_{2}$ and \protect$\theta=0$}
    \label{fig:BJJ-1: $U1=U3$}
\end{figure}

\begin{figure}[ht]
  \centering
  \subcaptionbox{Correlation matrix: \protect$<b_{i}^{\dagger}b_{j}>$}[.49\columnwidth]{%
    \includegraphics[width=\linewidth]{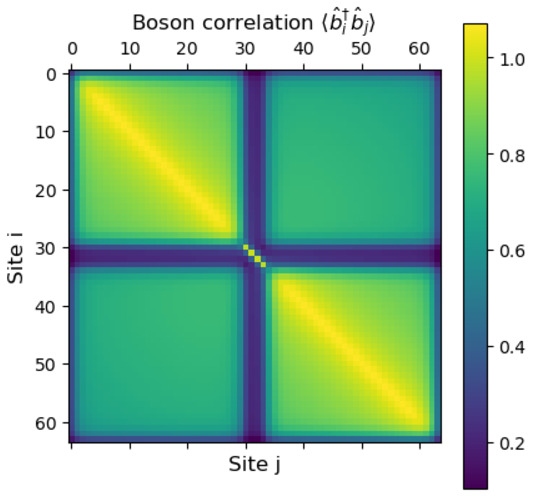}%
  }\hfill
  \subcaptionbox{Site occupation}[.49\columnwidth]{%
    \includegraphics[width=\linewidth]{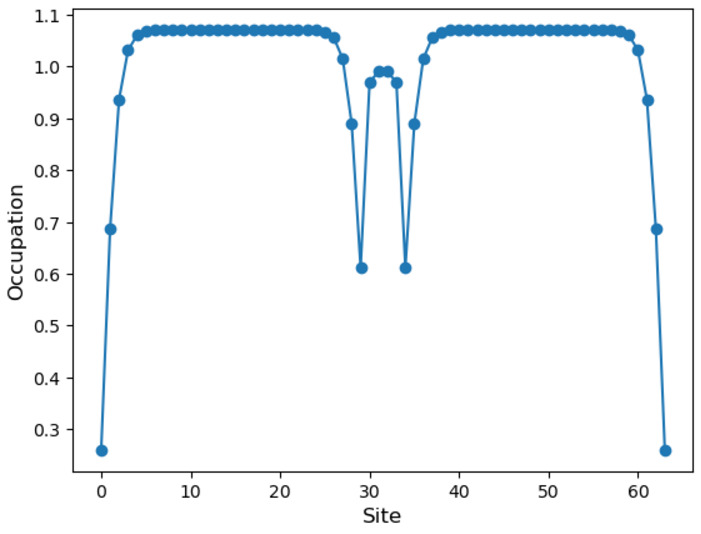}%
  }
  \caption{Type-1: Correlation matrix and density profile}
  \label{fig:type1 correlation matrix and density profile}
\end{figure}

The correlation matrix is divided into 9 regions in plot \ref{fig:type1 correlation matrix and density profile}. The top-right and bottom-left regions show the correlation between regions-1 and 3, which means these two regions are connected. Therefore, this indicates the possibility of tunneling of particles between the two superfluid regions when the ground-state is evolved. 
% This indicates the Josephson effect in this system. 
The dark regions in the middle indicate no correlations of region-2 with regions-1 and 3. The top-left and bottom-right regions indicate self-correlations within regions-1 and 3 respectively. The bright central region is the Mott-insulatior of region-2 where the $<b_{i}^{\dagger}b_{j}>$ becomes the number operator $<b_{i}^{\dagger}b_{i}>$. This bright central region indicates there is only one particle at each site (i.e, $<b_{i}^{\dagger}b_{i}> \approx 1$). 
% Therefore, this is one of the valid configurations as it's ground-state can show the Josephson effect under time evolution.

% In plot \ref{fig:type1 correlation matrix and density profile}, the top-left region contains the correlation matrix elements within region-1, followed by regions-1 and  2 and top-right includes the correlations between regions-1 and 3. The middle section from left is between regions-1 and 2, followed by correlations within region 2, and regions-2 and 3. A similar understanding can be applied the bottom regions in the correlation matrix. 

Site occupation observable also shows the symmetric nature of this configuration. Since this is a finite system, the wavefunction tends to vanish at the boundaries, which is why the probability of finding the particles at the boundary is almost equal to zero. The Mott-insulating region-2 has exactly one particle at each site.

\subsection{Types-2 and 3: \protect$U_{1} < U_{3} < U_{2}$ and \protect$U_{1} > U_{3} < U_{2}$}
\label{subsec type2 type3 BJJ}
\begin{figure}[ht]
  \centering
  \subcaptionbox{\protect$U_{1}=1, U_{3}=0.5, U_{2}=10$}[.49\columnwidth]{
    \includegraphics[width=\linewidth]{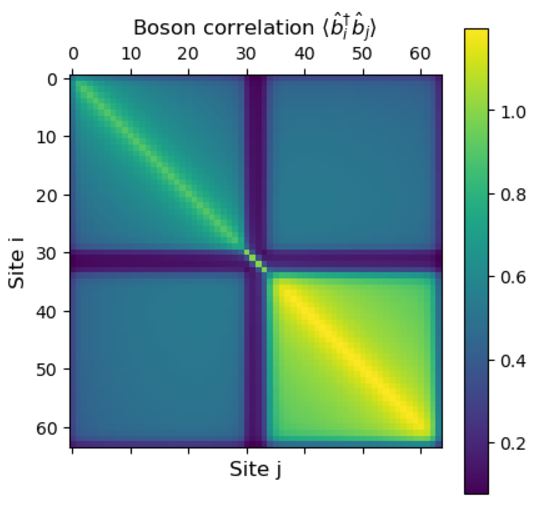}
  }\hfill
  \subcaptionbox{\protect$U_{1}=0.5, U_{3}=1, U_{2}=10$}[.49\columnwidth]{
    \includegraphics[width=\linewidth]{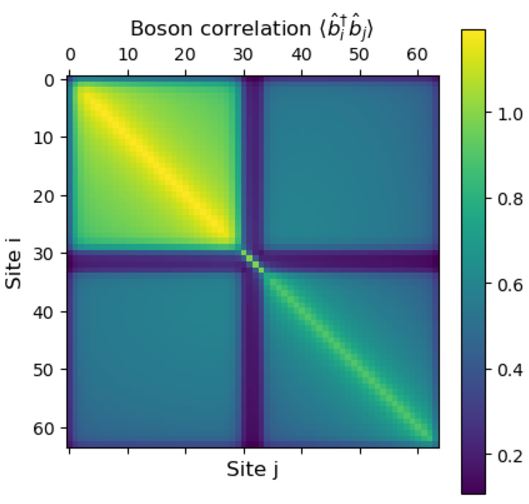}
  }
  \caption{Types-2 and 3: Correlation matrix}
  \label{fig:type2 and 3 correlation matrix}
\end{figure}

\begin{figure}[ht]
  \centering
  \subcaptionbox{\protect$U_{1}=1, U_{3}=0.5, U_{2}=10$}[.49\columnwidth]{
    \includegraphics[width=\linewidth]{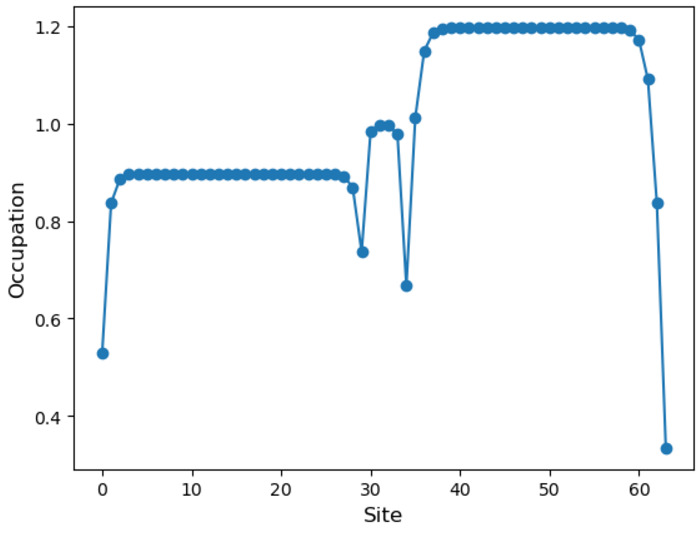}
  }\hfill
  \subcaptionbox{\protect$U_{1}=0.5, U_{3}=1, U_{2}=10$}[.49\columnwidth]{
    \includegraphics[width=\linewidth]{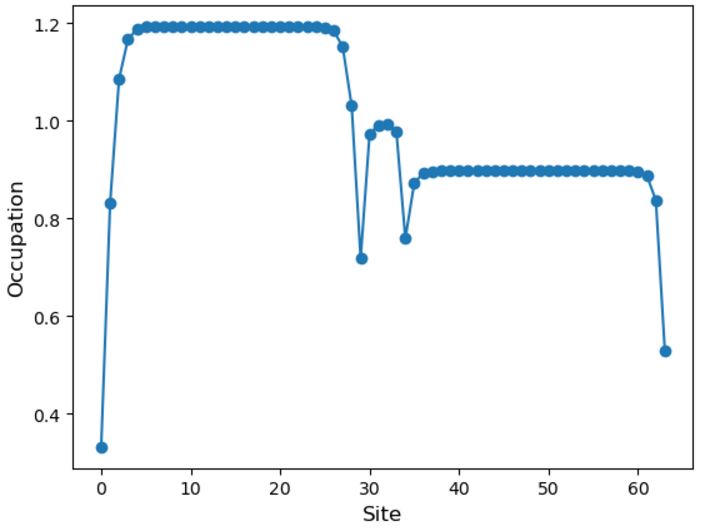}
  }
  \caption{Types-2 and 3: Site Occupation}
  \label{fig:type2 and 3 site occupation}
\end{figure}

In this configuration, $U_{1}\neq U_{3}$ introduces asymmetry in the system's eigenstates. Due to which, the particles tend to be positioned towards the direction in which the superfluid region with less value of on-site interaction potential as shown in \ref{fig:type2 and 3 correlation matrix}. Nevertheless, the two superfluid regions are connected, therefore, indicating the possibility of tunneling of particles across the regions-1 and 3 when this state is evolved. 
% Therefore, these configurations can be considered potentially valid configurations. 
These observations are also seen in the density profile for this system as shown in \ref{fig:type2 and 3 site occupation}. These plots are computed for finding the average density of particles at each site which is why there are fractional values too. And, the Mott-insulating region-2 has only one particle per site.

\subsection{Type-4: \protect$\theta_{1} = \theta_{3} < \theta_{2}$}
\label{subsec AJJ}
\begin{figure}[ht]
    \centering
    \includegraphics[width=0.8\linewidth]{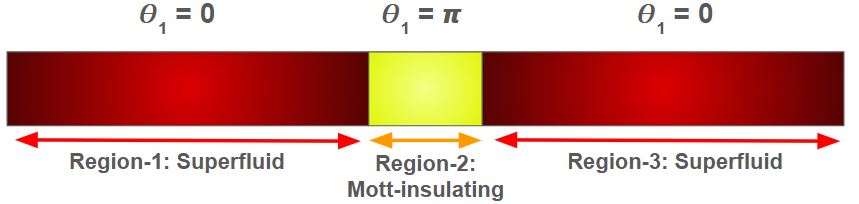}
    % \caption{$\theta_{1} = \theta_{3} < \theta_{2}$ and $U=0$}
    \caption{Anyonic Josephson junction: \protect$0$-$\pi$-$0$}
    \label{fig:AJJ-1: $theta1=theta3$}
\end{figure}

In this configuration, $\theta$ is varied while keeping $U = \{2, 0.5\}$ constant as shown in \ref{fig:AJJ-1: $theta1=theta3$}. $\theta_{1,3}=0$ results in a superfluid phase in regions-1 and 3, and $\theta_{2} = \pi$ results in an insulating phase generated by pseudofermions. 

\begin{figure*}[ht]
  \centering
  % First row of subfigures
  \begin{subfigure}[b]{0.23\textwidth}
    \includegraphics[width=\textwidth]{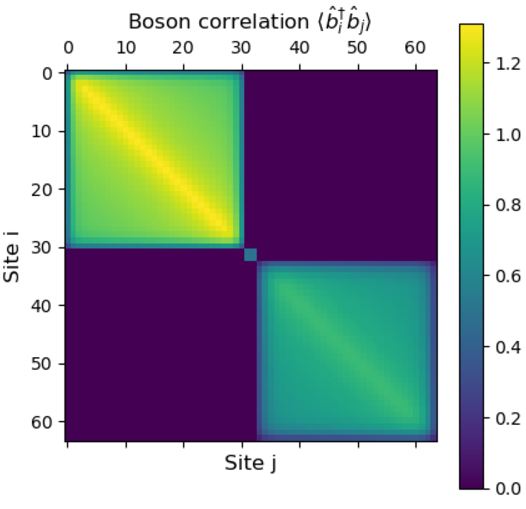}
    \caption{\protect$U=0.5$, \protect$N_{R_{2}, sites} =2$}
    \label{subfig1}
  \end{subfigure}
  \hfill
  \begin{subfigure}[b]{0.23\textwidth}
    \includegraphics[width=\textwidth]{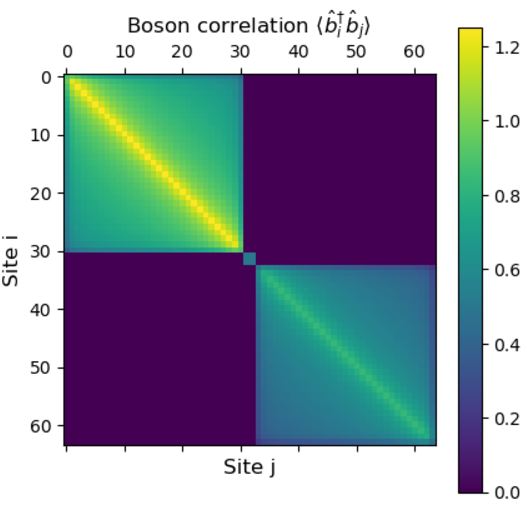}
    \caption{\protect$U=2$,  \protect$N_{R_{2}, sites} =2$}
    \label{subfig2}
  \end{subfigure}
  \hfill
  \begin{subfigure}[b]{0.23\textwidth}
    \includegraphics[width=\textwidth]{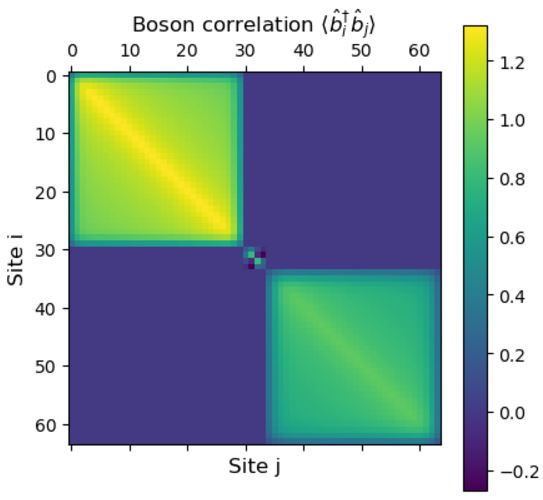}
    \caption{\protect$U=0.5$, \protect$N_{R_{2}, sites} =4$}
    \label{subfig3}
  \end{subfigure}
  \hfill
  \begin{subfigure}[b]{0.23\textwidth}
    \includegraphics[width=\textwidth]{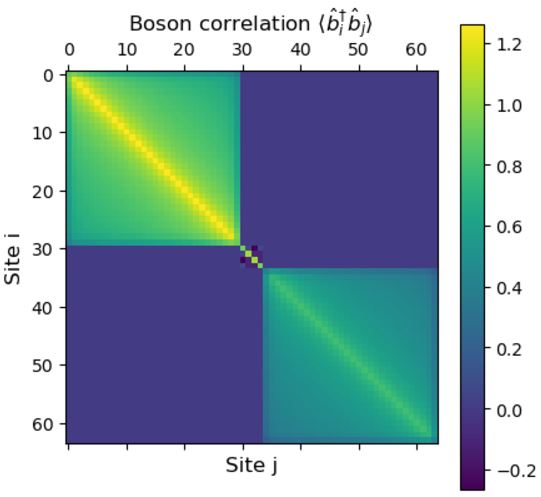}
    \caption{\protect$U=2$, \protect$N_{R_{2}, sites} =4$}
    \label{subfig4}
  \end{subfigure}
  
  \vspace{1em} % Adjust vertical space between rows
  
  % Second row of subfigures
  \begin{subfigure}[b]{0.23\textwidth}
    \includegraphics[width=\textwidth]{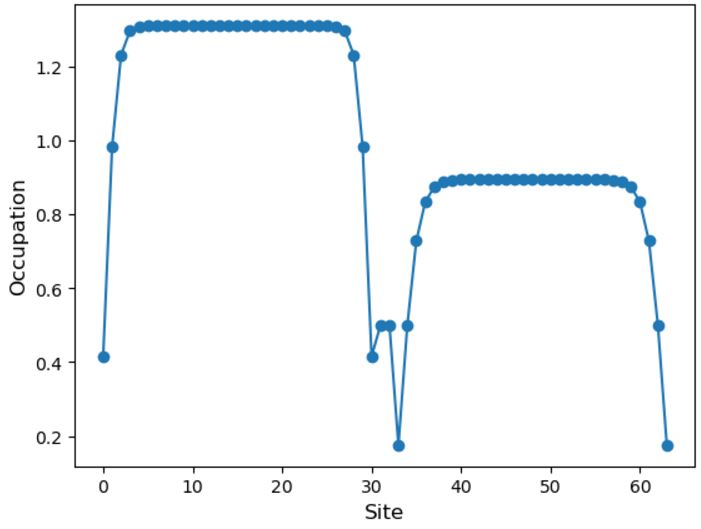}
    \caption{\protect$U=0.5$,  \protect$N_{R_{2}, sites} =2$}
    \label{subfig5}
  \end{subfigure}
  \hfill
  \begin{subfigure}[b]{0.23\textwidth}
    \includegraphics[width=\textwidth]{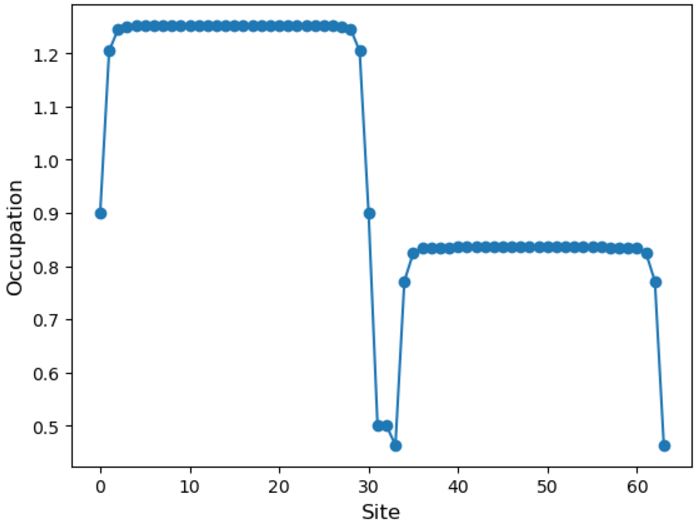}
    \caption{\protect$U=2$,  \protect$N_{R_{2}, sites} =2$}
    \label{subfig6}
  \end{subfigure}
  \hfill
  \begin{subfigure}[b]{0.23\textwidth}
    \includegraphics[width=\textwidth]{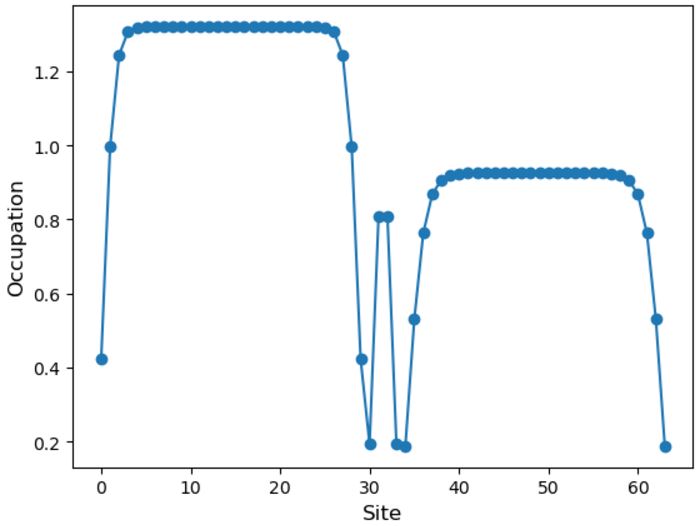}
    \caption{\protect$U=0.5$, \protect$N_{R_{2}, sites} = 4$}
    \label{subfig7}
  \end{subfigure}
  \hfill
  \begin{subfigure}[b]{0.23\textwidth}
    \includegraphics[width=\textwidth]{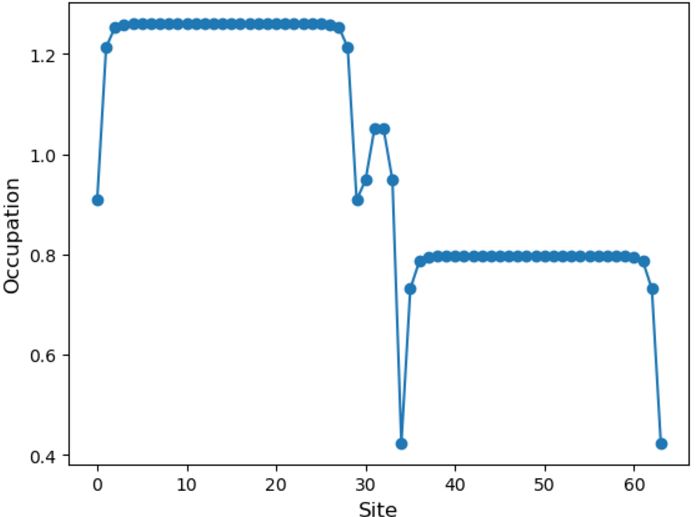}
    \caption{\protect$U=2$,  \protect$N_{R_{2}, sites} =4$}
    \label{subfig8}
  \end{subfigure}
  
  \caption{Correlation matrix and site occupation for \protect$U=\{0.5, 2\}$ and  \protect$N_{R_{2}, sites} =\{2,4\}$}
  \label{fig: AJJ - site occupation and correlation matrix for $U = 0.5, 2$ and Nr2 $sites = 2,4$}
\end{figure*}

The total system size for \ref{fig: AJJ - site occupation and correlation matrix for $U = 0.5, 2$ and Nr2 $sites = 2,4$} is 64 lattice sites with unit-filling case. The lattice sites in regions-1, 2 and 3 for \ref{subfig1}, \ref{subfig2}, \ref{subfig5}, \ref{subfig5} are 31, 2, 31 respectively while they are 30, 4, 30 for \ref{subfig3}, \ref{subfig4}, \ref{subfig7}, \ref{subfig8}.

From \ref{subfig1}, \ref{subfig2}, \ref{subfig3}, \ref{subfig4}, there are no correlations between regions-1 and 2 which might indicate that tunneling of particles cannot be possible. Therefore, the ground-state of this configuration of the model with $\theta_{1} = \theta_{3} = 0, \theta_{2} = \pi$, does not seem to produce the Josephson effect under time evolution as all the regions are disconnected. However, from the dynamics of this state as detailed in Section \ref{sec dynamics of AJJ}, we notice that there is tunneling of particles over time. This tells us that although the ground-state at time $t=0$ is a product state of the ground-states of all the three regions, however, when evolved with time, the product state becomes correlated over time.

% \FloatBarrier
% \FloatBarrier

In general for the type-4 anyonic Josephson junctions, correlations are stronger in regions with $U=0.5$ than with $U=2$. Correlations within region-1 are much more prominent than those within region-3 due to the left-bias introduced by the additional phase $e^{i\theta n_{j}}$ in the anyonic Hubbard model as given in \ref{boson_HamilAHM}. As a result, the particles have a tendency to be positioned towards the left-side of the system. This observation can be further validated using the site occupation of particles at each site. It would be interesting to study these observables in periodic boundary conditions to verify if this left-bias in the ground-state leads to symmetry in these ground-state observables about region-2 (i.e, having same values/ patterns in regions-1 and 3).

In each of the four different configurations of \ref{fig: AJJ - site occupation and correlation matrix for $U = 0.5, 2$ and Nr2 $sites = 2,4$}, the correlations and the density profile within region-2 do not show Mott-insulating behavior as in this phase for unit-filling case, it is expected to show one particle per site just as it is observed in \ref{fig:type1 correlation matrix and density profile}, \ref{fig:type2 and 3 correlation matrix} and \ref{fig:type2 and 3 site occupation}. There are correlations existing between the two sites in region-2 of \ref{subfig1} and \ref{subfig2} and site occupation in this case for these two sites are fractional and have the same values. 
% This may be due to bunching of bosons when kept on same site. When the bosons interact with each other strongly, they form a repulsive bound pair and act as a single entity and become localized. When they are weakly interacting with each other, this results in them performing random walks independent of each other as given in \cite{Preiss:2015tyr}. 

 In the case of region-2 with four lattice sites, there are anti-correlations observed within region-2, and the values of $<b_{i}^{\dagger}b_{i}>$ become more prominent and tend to one with increase in $U$. The anti-correlations can be attributed to the anti-bunching of pseudofermions when kept at adjacent sites, results in anti-correlations \cite{kwan2024realization}. And, $<b_{i}^{\dagger}b_{i}> \approx 1$ can be due to particles being localized due to fermionization. The central two sites of region-2 tend to be slightly more than one while the boundary sites of region-2 tend to be less than one. And, having anti-correlations along the off-diagonal elements of region-2 could mean increase in site occupancy in the central sites results in decrease in the site occupancy of the boundary sites. %% That is why we have fractional values at the boundaries compared to the bulk of region-2 in this case.

\section{Dynamics of Anyonic Josephson junctions}
\label{sec dynamics of AJJ}
To study the Josephson effect, the ground-state is changed such that the phase difference between two parts of each configuration is $\phi$. The application of the phase difference operator $ph_{Op} = e^{i\phi \hat{N}}$ on the ground-state can be done in two ways through - (i) symmetrical and (ii) asymmetrical ways. 

In the symmetrical application of the phase difference operator, it is applied such that the phase difference between two regions is $\phi$ in which part-1, including region-1 and half of region-2 has a phase $\phi$ while part-2 has phase $0$. The two parts are constructed this way to maintain symmetry in the system which will make the analysis convenient. Here, $\hat{N}$ is number operator of sites in regions-1 and half of region-2. 

While in the asymmetrical application, it is applied between two uneven partition of the lattice sites. For example, if the lattice system of four sites, then $\hat{N}$ is the number operator of the first three sites, so the form of this number operator would be $\hat{N} = \hat{n}_{1}\otimes\hat{n}_{2}\otimes\hat{n}_{3}\otimes\hat{I}_{4}$ where $\hat{I}$ is the identity operator and $\hat{n}_{i}$ is number operator at site $i$. All the simulations in this section are done using exact diagonalization unless mentioned otherwise, and the phase difference takes the values of $\phi \in \{\pi, \frac{\pi}{4}\}$.

For our analysis, we use dynamical observable quantities - (i) rate of change of particle number difference between two regions of the system (or the population imbalance as a function of time \cite{brollo2022two}) and is given by $z = \frac{<\hat{N}_{1}>-<\hat{N}_{2}>}{N}$ where $N$ is the total number of particles in the system, (ii) density profile at each instant of time, and (iii) Correlation matrix elements at each instant of time, including the magnitude and complex phase of each matrix element as a function of time.

\subsection{Two-site anyonic Josephson junction}
\label{subsec two site AJJ} 
\begin{figure}[ht]
  \centering
  \subcaptionbox{$\phi = \pi, U = 0$\label{subfig:U = 0, pi=0, two site model}}
  [.5\columnwidth]{
    \includegraphics[width=\linewidth]{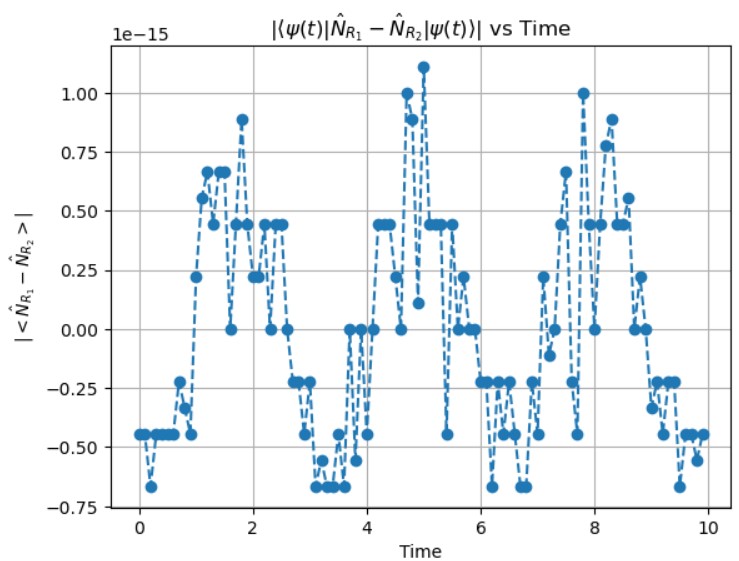}
  }\hfill
  \subcaptionbox{$\phi = \frac{\pi}{4}, U=0$\label{subfig:U = 0, pi/4, two site model}}
  [.5\columnwidth]{
    \includegraphics[width=\linewidth]{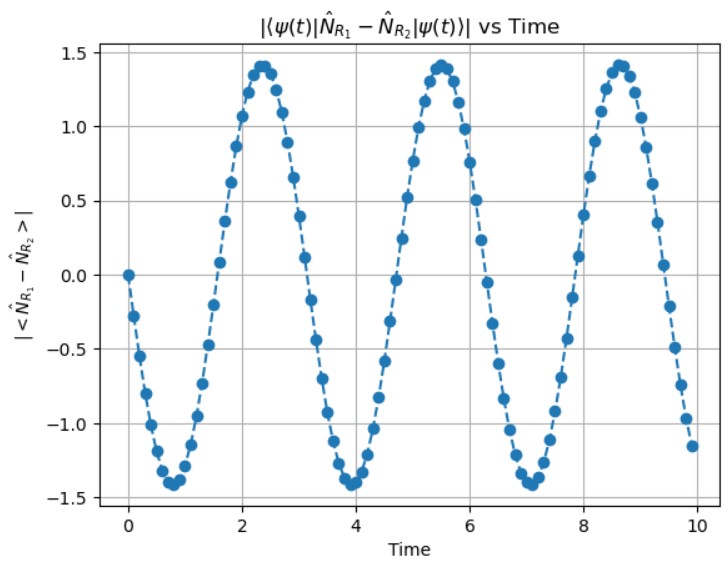}
  }
  % \hfill
  \subcaptionbox{$\phi = \frac{\pi}{4}, U=0.5$\label{subfig:U = 0.5, pi/4, two site model}}
  [.8\columnwidth]{
    \includegraphics[width=\linewidth]{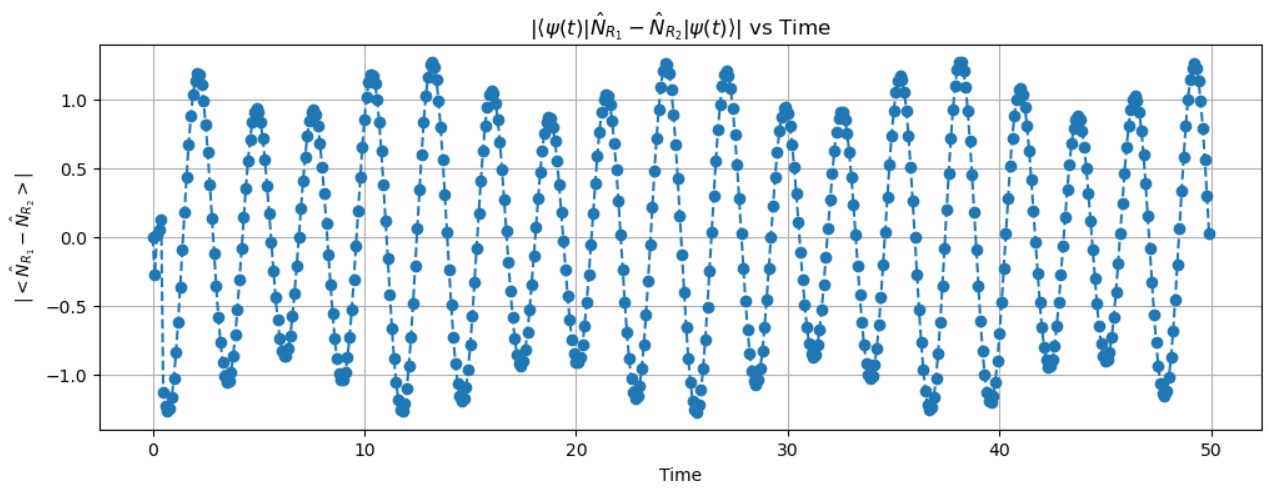}
  }
  \caption{Rate of change of particle number difference in two-site model}
  \label{fig:Rate of change of particle number in two-site model}
\end{figure}

% For the multiples of $\phi = \pi$ including zero phase difference, there is no net particle difference between the two superfluid regions, over time. 
For a two-site anyonic Josephson junction, it is evident from plot \ref{fig:Rate of change of particle number in two-site model} that the rate of change of particle (or population) difference between two sites is zero for $\phi = \pi$ irrespective of the values of $U$ \footnote{It was studied for $U={0, 0.5, 1, 1.5}$ and $J=1$ is constant}.

\begin{figure*}[ht]
  \centering
  % First row of subfigures
  \begin{subfigure}[b]{0.23\textwidth}
    \includegraphics[width=\textwidth]{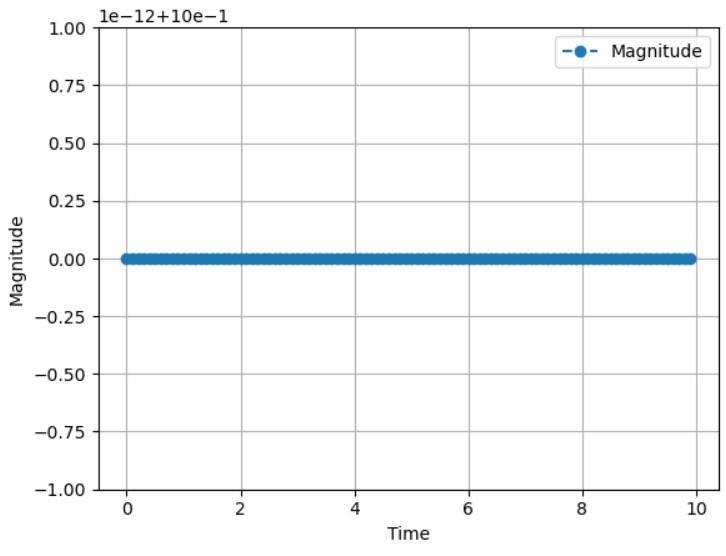}
    \caption{$Real(b^{\dagger}_{1}b_{1})$}
    \label{subfig1: b11 real}
  \end{subfigure}
  \hfill
  \begin{subfigure}[b]{0.23\textwidth}
    \includegraphics[width=\textwidth]{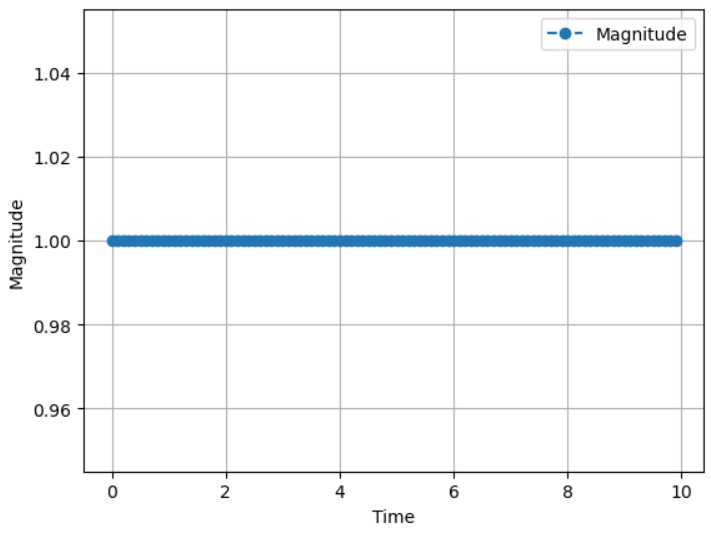}
    \caption{$Real(b^{\dagger}_{1}b_{2})$}
    \label{subfig2 - real part of bb12}
  \end{subfigure}
  \hfill
  \begin{subfigure}[b]{0.23\textwidth}
    \includegraphics[width=\textwidth]{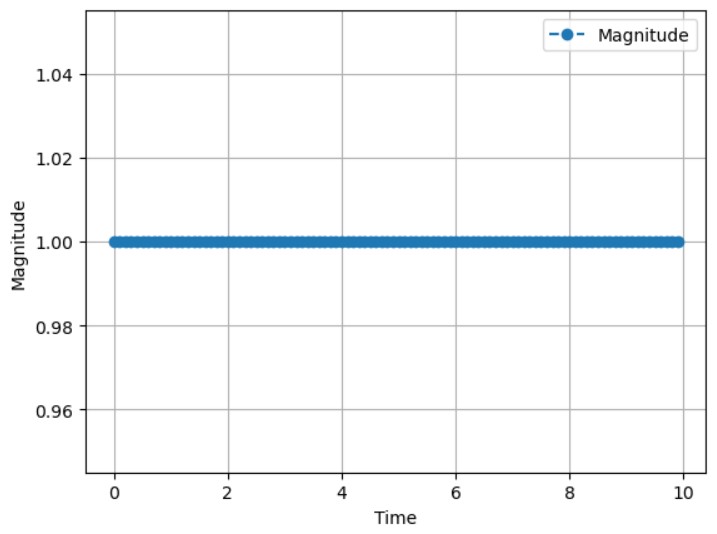}
    \caption{$Real(b^{\dagger}_{2}b_{1})$}
    \label{subfig3 - real part of bb21}
  \end{subfigure}
  \hfill
  \begin{subfigure}[b]{0.23\textwidth}
    \includegraphics[width=\textwidth]{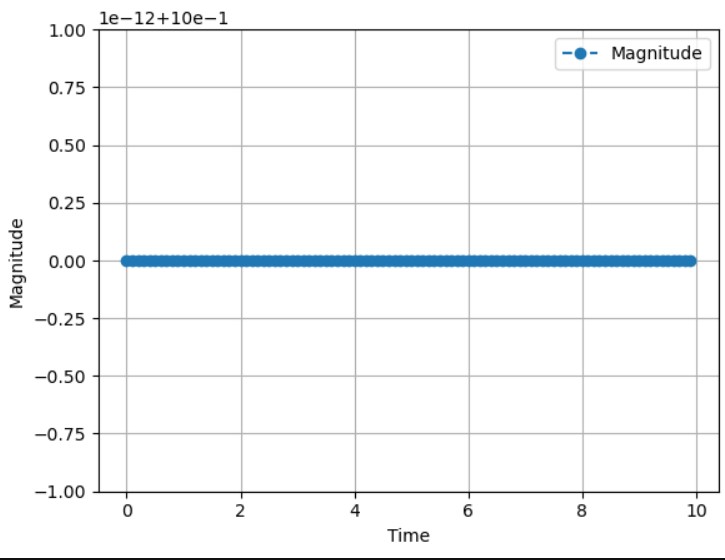}
    \caption{$Real(b^{\dagger}_{2}b_{2})$}
    \label{subfig4 - real part of bb22}
  \end{subfigure}
  
  \vspace{1em} % Adjust vertical space between rows
  
  % Second row of subfigures
  \begin{subfigure}[b]{0.23\textwidth}
    \includegraphics[width=\textwidth]{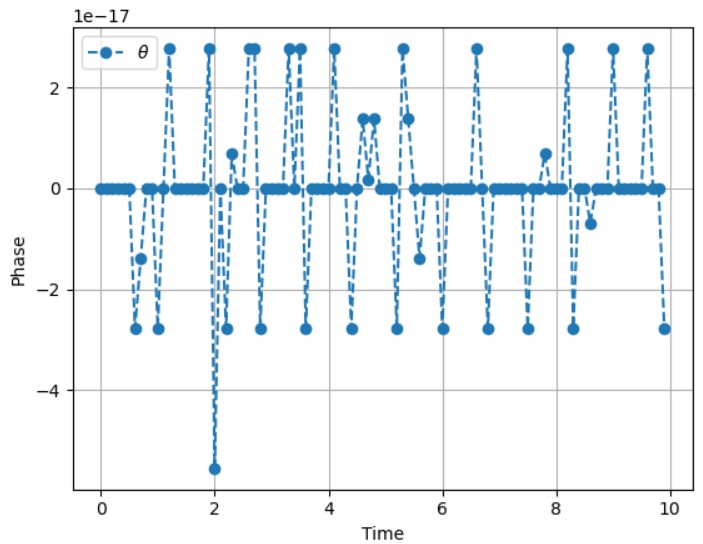}
    \caption{$Im(b^{\dagger}_{1}b_{1})$}
    \label{subfig5 - imag of bb11}
  \end{subfigure}
  \hfill
  \begin{subfigure}[b]{0.23\textwidth}
    \includegraphics[width=\textwidth]{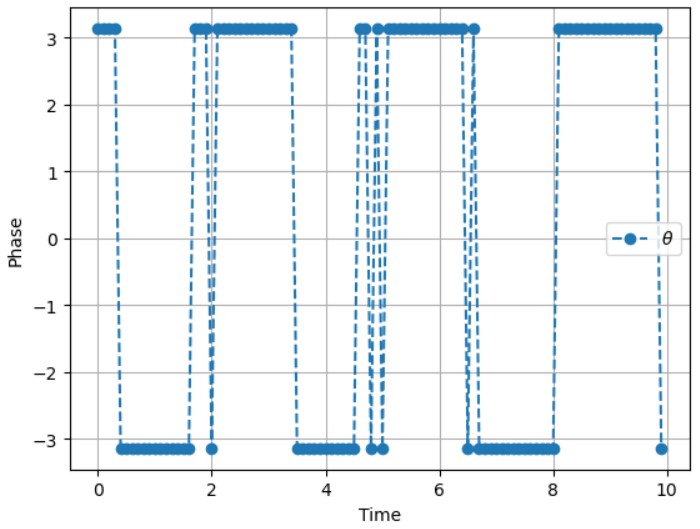}
    \caption{$Im(b^{\dagger}_{1}b_{2})$}
    \label{subfig6 - imag part of bb12}
  \end{subfigure}
  \hfill
  \begin{subfigure}[b]{0.23\textwidth}
    \includegraphics[width=\textwidth]{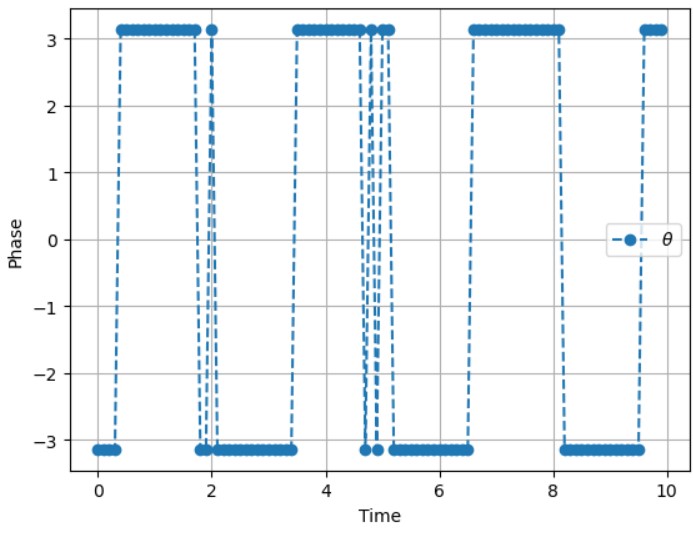}
    \caption{$Im(b^{\dagger}_{2}b_{1})$}
    \label{subfig7 - imag part of bb21}
  \end{subfigure}
  \hfill
  \begin{subfigure}[b]{0.23\textwidth}
    \includegraphics[width=\textwidth]{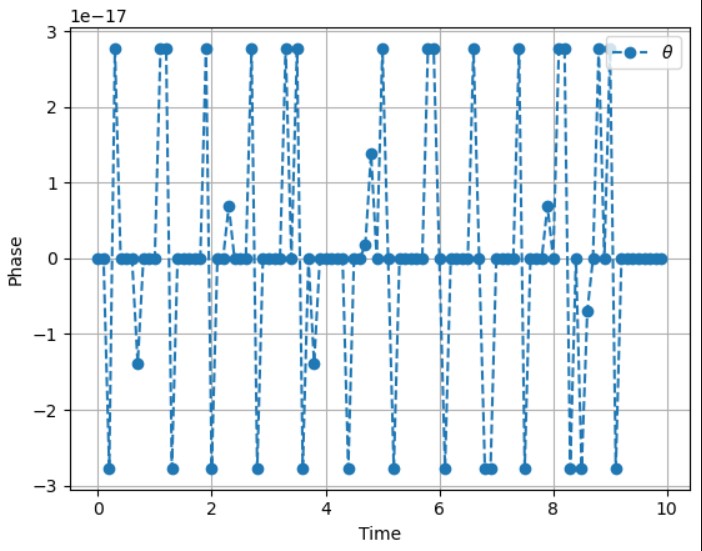}
    \caption{$Im(b^{\dagger}_{2}b_{2})$}
    \label{subfig8 - image of bb22}
  \end{subfigure}
  
  \caption{Real and imaginary parts of each correlation matrix element as a function of time for the two-site model with \protect$U=0$ and \protect$\phi = \pi$. The first (second) row shows the real (imaginary) parts of each correlation matrix element.}
  \label{fig: Real and imaginary parts of each correlation matrix elements for U0 and phi_pi for the two-site model}
\end{figure*}

This is further validated from the plots of correlation matrix diagonal elements as a function of time, we notice that they have constant values (i.e one, in this case as we are studying for the unit-filling case) in their real parts. And, their imaginary parts have zero values. This is because the diagonal elements of the correlation matrix are the number operators at each site, i.e, $b^{\dagger}_{1}b_{1} = n_{1}$ and $b^{\dagger}_{2}b_{2} = n_{2}$. 

However, for the non-diagonal elements of the correlation matrix, their real parts do not vary with time while the values of their imaginary parts oscillate between $\pi$ and $-\pi$, but since the phase difference in the state is the same, the $\pm$ signs do not matter. When the phase difference between the two sites is $\pi$, the density profile of this state is constant with time, and the distribution is symmetric.

For the case of $\phi = \frac{\pi}{4}$ and $U = 0$, the population difference between the two sites as a function of time takes non-zero finite values, and the flow of particles is oscillatory. This is also verified from its correlation matrix elements that change with time. These observations match qualitatively with the mean field theory analysis of \cite{brollo2022two} for the two-site model with $\theta = 0$. 

For $\phi = \frac{\pi}{4}$ with $U\neq 0$ forms a periodic pattern in which the amplitude of particle number difference changes over time. And, this change in amplitude is periodic in nature as shown in \ref{subfig:U = 0.5, pi/4, two site model}. While mean-field analysis of this model with $\phi = \frac{\pi}{4}$ and $U\neq 0$ predicts decaying oscillations \cite{brollo2022two}, an exact treatment in our analysis reveals sustained oscillations. This discrepancy arises because the exact analysis captures the quantum effects more accurately, that are neglected in the mean-field approximation.

% \begin{figure}[ht]
%     \centering
%     \includegraphics[width=1\linewidth]{images/two-site_pics/nr1_nr2_Piby4_U0.5.PNG}
%     \caption{Rate of change of particle number in two-site model with $U=0.5, \phi=\frac{\pi}{4}$}
%     \label{fig: Rate of change of particle number in two-site model with U = 0.5 and phi = pi/4}
% \end{figure}

\begin{figure}[ht]
  \centering
  \subcaptionbox{$\phi = \pi, U = 0$\label{subfig:U = 0, pi=0, six site model}}
  [.5\columnwidth]{
    \includegraphics[width=\linewidth]{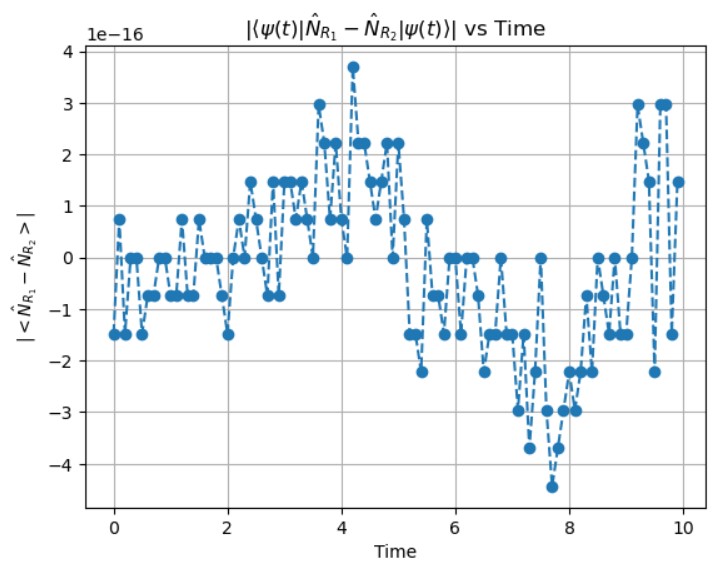}
  }\hfill
  \subcaptionbox{$\phi = \frac{\pi}{4}, U=0$\label{subfig:U = 0, pi/4, six site model}}
  [.5\columnwidth]{
    \includegraphics[width=\linewidth]{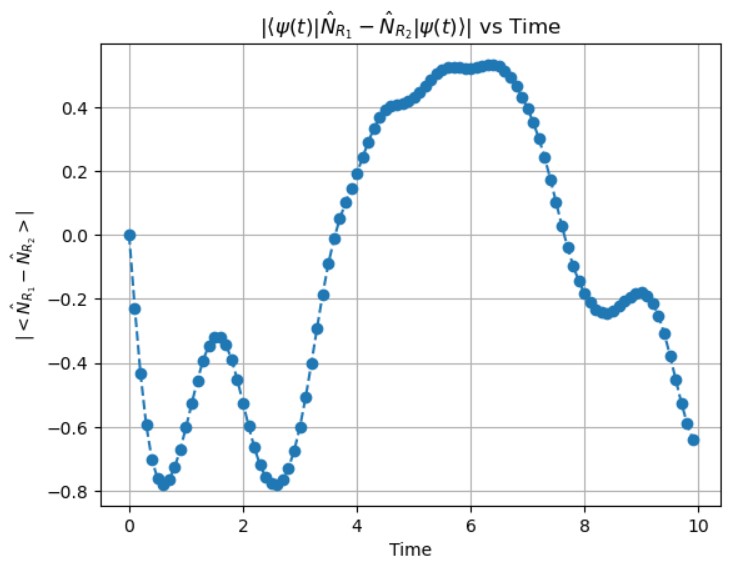}
  }
  % \hfill
  \subcaptionbox{$\phi = \frac{\pi}{4}, U=1.5$\label{subfig:U = 1.5, pi/4, six site model}}
  [.8\columnwidth]{
    \includegraphics[width=\linewidth]{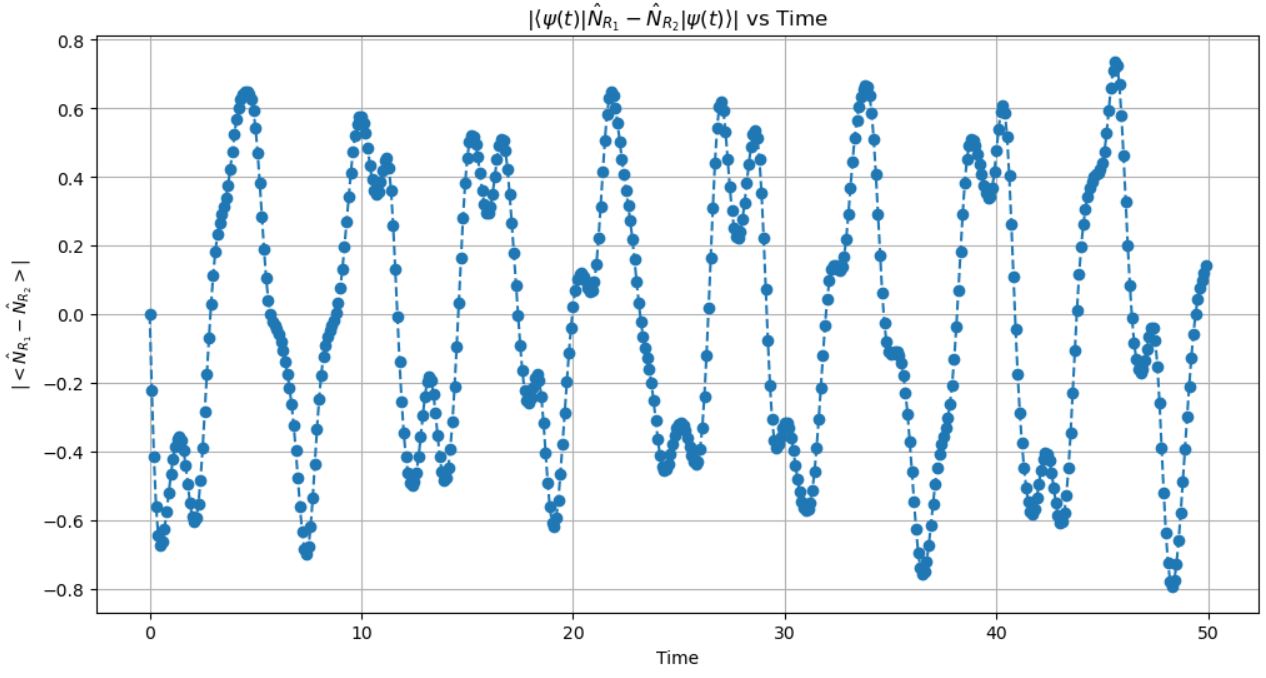}
  }
  \caption{Rate of change of particle number difference in a 6 lattice sites model with only two superfluid regions (without any insulating barrier) having phase difference $\phi={0, \frac{\pi}{4}}$ under symmetrical application  of the phase difference operator.}
  \label{fig:Rate of change of particle number in six lattice model}
\end{figure}

From this analysis, we can conclude that the two-site model with $\theta = 0$ using bosons, seems to show properties similar to the Josephson effect. Since this is in agreement with the supercurrent density $J_{s} = J_{o}\sin{\phi}$ in the Josephson junctions where $J_{s} = 0$ for phase differences that are multiples of $\pi$ and $J_{s}$ is non-zero for other values of $\phi$. Also, note that in the two-site model, we can only apply the phase difference operator symmetrically while the asymmetrical application would be irrelevant for our analysis in the two-site model. 

\begin{figure}[ht]
  \centering

  \subcaptionbox{$N_{sites}<N_{particles}$\label{subfig:U1=U3=0,U3=10, pi/4, six site model N = 6, smaller than k = 7}}
  [.45\columnwidth]{
    \includegraphics[width=\linewidth]{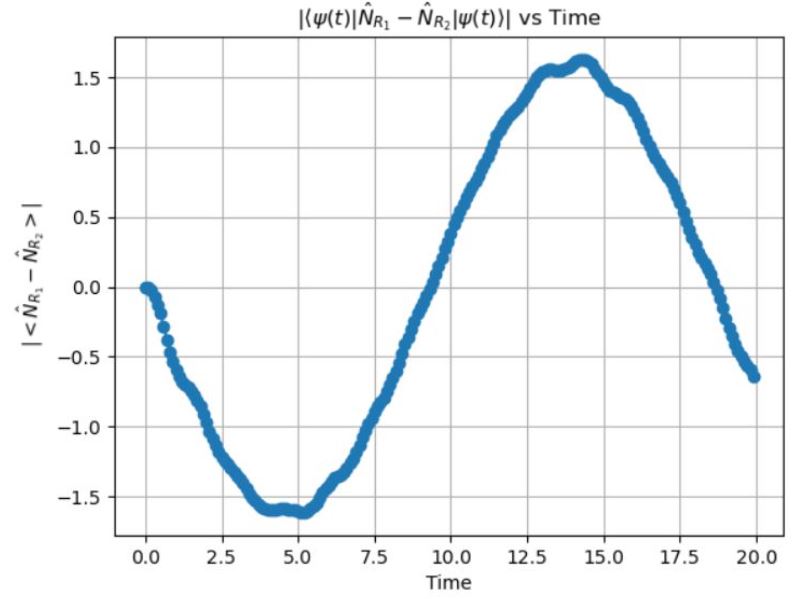}
  }\hfill
  \subcaptionbox{$N_{sites}=N_{particles}$\label{subfig:U1=U3=0,U3=10, pi/4, six site model unit filling}}
  [.45\columnwidth]{
    \includegraphics[width=\linewidth]{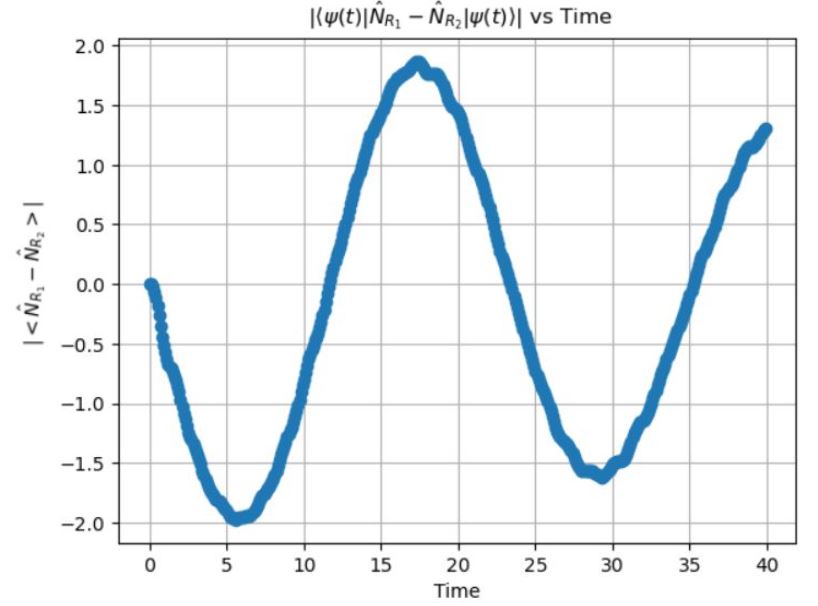}
  }
  \caption{Rate of change of particle number difference in a type-1 anyonic Josephson junction with 6 lattice sites ($N_{sites, R_{1,2,3}}= 2$), $\phi=\frac{\pi}{4}, U_{1,3}=0, U_{2}=10$ under symmetrical application of phase difference operator.}
  \label{fig:Rate of change of particle number in type-1 BJJ six lattice model for different particle fillings}
\end{figure}

\subsection{For more than two-sites}
\label{subsec: for more than two sites}
For more than two-sites, there are three types of models that are considered here (i) only two superfluid regions not separated by any insulating region (i.e the entire lattice system will be in superfluid phase); (ii) type-1 anyonic Josephson junction configurations; and (iii) type-4 anyonic Josephson junction. 

\begin{figure}[ht]
  \centering

  \subcaptionbox{$\phi = \pi$\label{subfig:theta1=theta3=0,theta2=pi, pi, six site model N = 6 = k}}
  [.45\columnwidth]{
    \includegraphics[width=\linewidth]{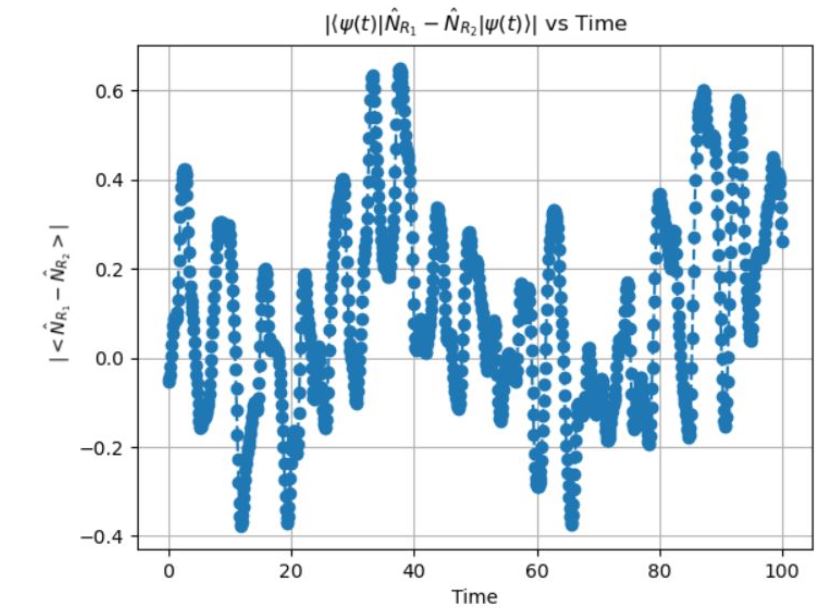}
  }\hfill
  \subcaptionbox{$\phi = \frac{\pi}{4}$\label{ssubfig:theta1=theta3=0,theta2=pi, pi/4, six site model N = 6 = k}}
  [.45\columnwidth]{
    \includegraphics[width=\linewidth]{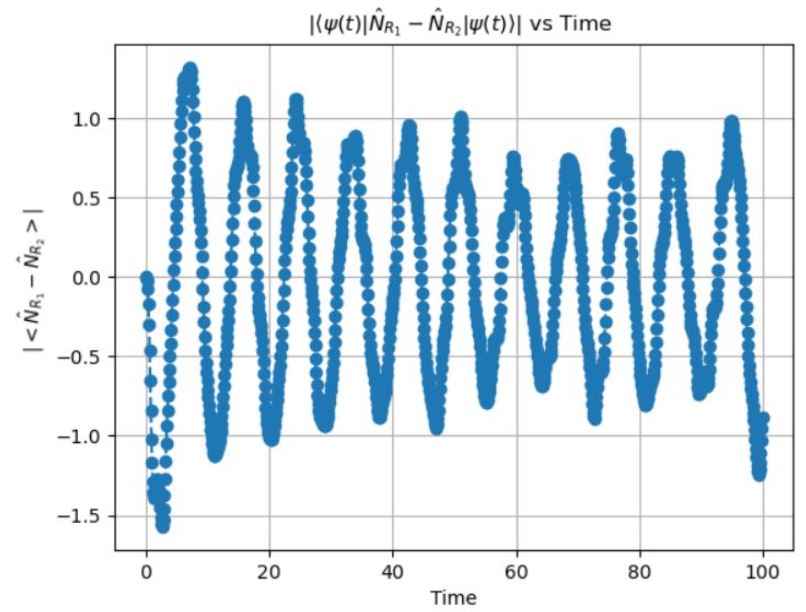}
  }
  \caption{Particle number difference vs time in a type-4 anyonic Josephson junction with 6 lattice sites ($N_{sites, R_{1,2,3}}= 2$), $\phi=\{\pi,\frac{\pi}{4}\}, \theta_{1,3}=0, \theta_{2}=\pi, U=0.5$. Time evolutions upto $T = 100$.}
  \label{fig:Rate of change of particle number in type-4 AJJ six lattice model}
\end{figure}

For the first two types (i) only two superfluid regions not separated by any insulating region; and (ii) type-1 anyonic Josephson junction configurations, the following conclusions are applicable. It is observed that the net particle number difference across the two regions is zero for $\phi = \pi$ for even number of lattice sites, although the correlations between particles on different sites do not remain constant and the change in particle number at each site is evident in their density profile. Since the site occupancy has symmetric distribution, i.e, the total number of particles between the two regions remains the same, this leads to the net particle number difference being zero. These observations are valid when the phase difference operator is applied symmetrically.

For system with odd number of lattice sites, the particle density distribution is not symmetric, therefore the net change in the particle number across the two regions is non-zero. Thus, for $\phi = \pi$, there is particle flow between the two regions as it is evident from the correlation matrix elements. These observations are valid for even number of lattice sites under asymmetrical application of the phase difference operator. In a system with odd number of lattice sites, only asymmetrical application of the phase difference operator is possible. 

% For $\phi=\frac{\pi}{4}$, there exists a net change in particle number difference between the two regions with time. Site occupancy does not have a symmetric distribution or shows a periodic pattern unlike the case of the two-site model. It will be interesting to carry out the simulations to study these dynamics in large system sizes in order to verify if these observations are also influenced by the finite-size effects.

%%%% This is true for asymmetry and symmetry application of phase difference operator.
For $\phi=\frac{\pi}{4}$, the particle number difference between the two regions evolves over time without a pattern, in contrast to the periodic behavior observed in the two-site model. Unlike the two-site case, the site occupancy does not exhibit a symmetric distribution. To assess whether these observations are affected by finite-size effects, it would be useful to extend the simulations to larger system sizes and analyze the resulting dynamics.

\begin{figure}[ht]
  \centering
  \subcaptionbox{Correlation matrix elements at time $T=2$\label{subfig:BJJ 64 lattice - correl  matrix}}
  [.45\columnwidth]{
    \includegraphics[width=\linewidth]{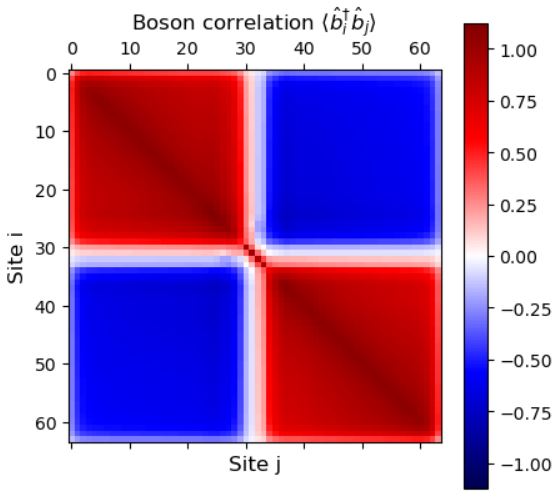}
  }\hfill
  \subcaptionbox{Density distribution at time $T=2$\label{subfig:BJJ 64 lattice - density distribution}}
  [.45\columnwidth]{
    \includegraphics[width=\linewidth]{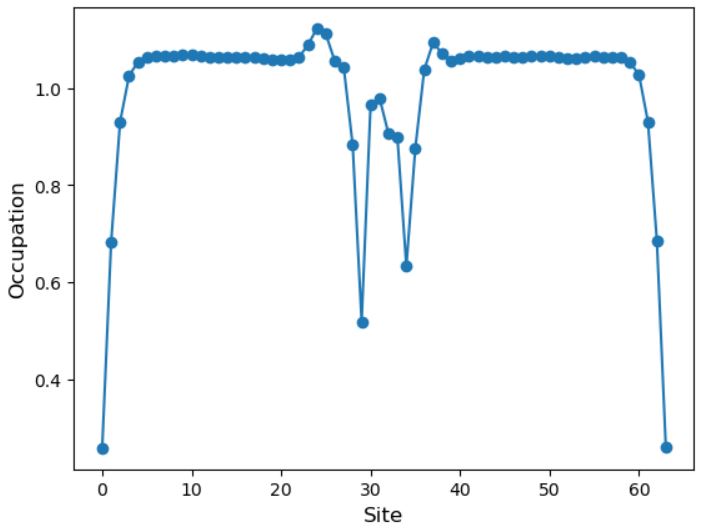}
  }
  % \hfill
  \subcaptionbox{Population imbalance as a function of time upto $T=2$\label{subfig:BJJ 64 lattice - population imbalance}}
  [.45\columnwidth]{
    \includegraphics[width=\linewidth]{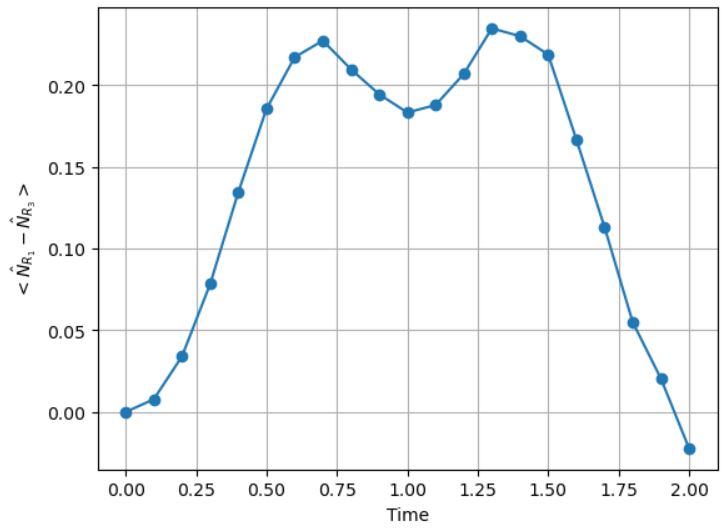}
  }
  \caption{Dynamical observables of 64 lattice sites of type-1 anyonic Josephson junction with $\phi=\pi, U_{1,3} = 0.5, U_{2} = 10, J=1$ using DMRG simulations under asymmetrical application of the phase operator. The changes in the correlation matrix elements seem to be extremely small, and in that of its density distribution happen only near the boundaries between the insulating and the superconducting regions, and within the insulating region.}
  \label{fig:BJJ dynamics in 64 lattice model}
\end{figure}

On analyzing the dynamics of type-4 anyonic Josephson junction configuration with $U=0.5$ irrespective of application of $ph_{Op}$ symmetrically or asymmetrically, it is observed that for $\phi = \{\pi, \frac{\pi}{4}\}$, the particle number difference between the two equal parts of the lattice has a finite non-zero value. For $\phi = \frac{\pi}{4}$ seems to show sustained oscillations with periodicity similar to the two-site model with $U\neq 0$.

Furthermore, from our analysis above for various configurations under open boundary conditions, it is evident that there is continuous flow of particles over time, i.e continuous current flow, in the anyonic Josephson junction model without any external biasing. This happens even when the particles are interacting with each other (here, in our setup, we have only considered nearest neighbor interactions and on-site interaction). This continual current flow is generated solely by producing an initial phase difference within the system. 

% Therefore, it would be interesting to realize the anyonic Josephson junction experimentally in optical systems (based on the proposals of \cite{Keilmann:2010cm, greschner2015anyon}) and benchmark with the conventional Josephson junctions. As this may potentially open avenues where optically developed anyonic Josephson junctions can replace conventional Josephson junctions in technologies such as precision technologies and quantum hardware systems.

Our numerical results indicate that these phenomena experimentally realized in optical platforms (based on the proposals of \cite{Keilmann:2010cm, greschner2015anyon}), could create tunable analogs of superconducting Josephson junctions. Such systems would provide a novel testbed for investigating many-body quantum effects and could enable the development of anyon-based quantum technologies. Future work could benchmark these systems against conventional Josephson junctions, which may further reveal potential applications in quantum simulation and high-precision measurement devices.

\section{Conclusions}
The type-4 anyonic Josephson junction configuration in which the insulating region is formed by the pseudofermions results in disconnected regions in its ground-state. In its insulating region, anti-correlations are observed as a result of anti-bunching of the pseudofermions. 

For the two-site model, a phase difference of $\pi$ involves no particle flow across the two sites. And, for $\frac{\pi}{4}$ phase difference, particle number difference as a function of time shows a periodic pattern in which $U=0$ oscillates with a constant amplitude while for $U\neq 0$ oscillates with a varying amplitude however, this variation in amplitude is also periodic.

For more than two sites, there exists particle flow across two regions for phase difference of $\pi, \frac{\pi}{4}$. However, under the symmetrical application of phase difference operator, the net change in particle number between the two regions is zero only for even number of lattice sites as density profile is symmetric and a phase difference $\pi$. Under asymmetrical application of phase difference operator and for odd number of lattice sites, there is always particle flow irrespective of the values of phase difference. These observations are valid for (i) type-1 anyonic Josephson junction and (ii) 1D lattice in superfluid phase. 

For type-4 anyonic Josephson junction with a phase difference of $\frac{\pi}{4}$, the particle number difference as a function of time results in decaying oscillations; and has a non-zero finite value for $\pi$.

% \section*{Acknowledgements}
% I acknowledge Professor Shovan Dutta's valuable discussion on this. I am grateful to Alessandro Veronese to clear some of my queries on their paper \cite{brollo2022two}.

\section*{Data Availability}
% All the codes including results and data of simulations can be found in \cite{Jessica_John_Britto_AnyonicJosephsonJunction}
The data that support the findings of this study will
be openly available following an embargo at the following
URL/DOI: http://bit.ly/4eezIVw \cite{Jessica_John_Britto_AnyonicJosephsonJunction}.

% \section*{References}

% \clearpage
% \onecolumn
% \onecolumngrid
\appendix
% \pagenumbering{alph}
\section{Mean Field Analysis of the two-site model}
\label{sec: MFT}
\begin{figure}[ht]
    \centering
    \includegraphics[width=0.5\linewidth]{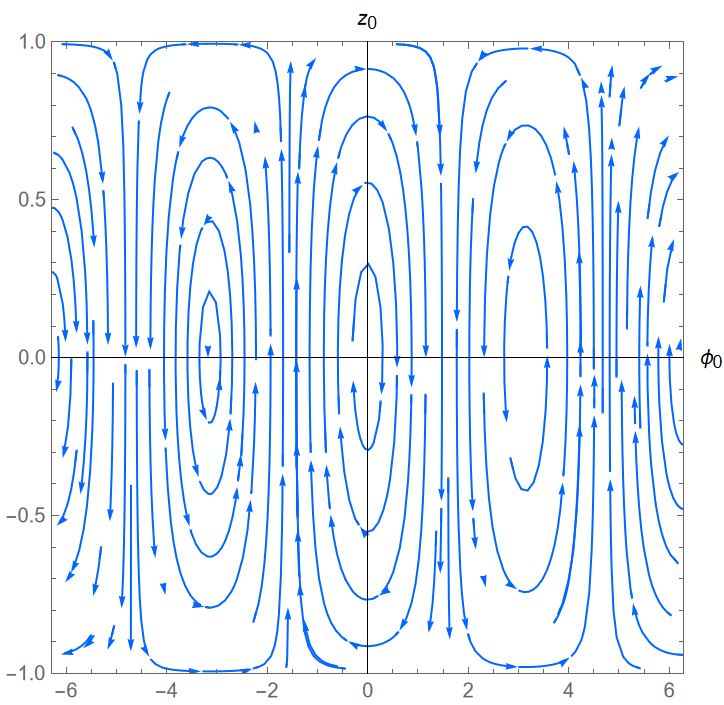}
    \caption{Phase portrait of $\phi$ and $z$ using the ODEs for the two-site model.}
    \label{fig:phase portrait using ODEs phi and z}
\end{figure}
This section briefly summarizes the key points from \cite{brollo2022two} to understand the phase portrait given in \ref{fig:phase portrait using ODEs phi and z} in which coherent Glauber states are used (whereas, in the exact treatment, we use the Fock states). The ordinary differentials equations are derived starting from the Hamiltonian of the two-site model which are then transformed into it's Lagrangian. From the Lagrangian, we obtain the following set of equations when $\theta = 0$.
\begin{equation}
\begin{split}
       \hbar \dot{\phi} & = \frac{Jz}{\sqrt{1-z^{2}}}\cos{\phi} + \frac{NUz}{4} \\
    \hbar \dot{z} & = -J\sqrt{1-z^{2}}\sin{\phi}
\end{split}
\end{equation}
while for the numerical simulations, $\hbar$ is taken to be one for simplicity. From the phase portrait, it is clear that $(\phi_{o}, z_{o}) = (m\pi, 0)$ is a stable point where $m = 0, 1, 2, ..$.

\section*{References}
\bibliographystyle{iopart-num}
\nocite{*}
% \bibliography{sample}
\bibliography{main}

\providecommand{\newblock}{}
\begin{thebibliography}{10}
\expandafter\ifx\csname url\endcsname\relax
  \def\url#1{{\tt #1}}\fi
\expandafter\ifx\csname urlprefix\endcsname\relax\def\urlprefix{URL }\fi
\providecommand{\eprint}[2][]{\url{#2}}

\bibitem{Keilmann:2010cm}
Keilmann T, Lanzmich S, McCulloch I and Roncaglia M 2011 {\em Nature Commun.\/} {\bf 2} 361 (\textit{Preprint} \eprint{1009.2036})

\bibitem{greschner2015anyon}
Greschner S and Santos L 2015 {\em Physical review letters\/} {\bf 115} 053002

\bibitem{kwan2024realization}
Kwan J, Segura P, Li Y, Kim S, Gorshkov A~V, Eckardt A, Bakkali-Hassani B and Greiner M 2024 {\em Science\/} {\bf 386} 1055--1060

\bibitem{brollo2022two}
Brollo A, Veronese A and Salasnich L 2022 {\em Physical Review A\/} {\bf 106} 023308

\bibitem{lerda2008anyons}
Lerda A 2008 {\em Anyons: Quantum mechanics of particles with fractional statistics\/} vol~14 (Springer Science \& Business Media)

\bibitem{zhai2023block2}
Zhai H, Larsson H~R, Lee S, Cui Z~H, Zhu T, Sun C, Peng L, Peng R, Liao K, Tölle J, Yang J, Li S and Chan G~K~L 2023 {\em The Journal of Chemical Physics\/} {\bf 159} 234801 ISSN 0021-9606

\bibitem{Jessica_John_Britto_AnyonicJosephsonJunction}
John~Britto J 2024 {Anyonic Josephson Junction} \url{https://github.com/JessicaJohnBritto/MTP}

\bibitem{Preiss:2015tyr}
Preiss P~M, Ma R, Tai M~E, Lukin A, Rispoli M, Zupancic P, Lahini Y, Islam R and Greiner M 2015 {\em Science\/} {\bf 347} 1260364

\bibitem{Lau:2020enu}
Lau L~L~H and Dutta S 2022 {\em Phys. Rev. Res.\/} {\bf 4} L012007 (\textit{Preprint} \eprint{2012.03977})

\bibitem{Freericks:1993gq}
Freericks J~K and Monien H 1993 {\em IOP\/}

\bibitem{greiner2002quantum}
Greiner M, Mandel O, Esslinger T, H{\"a}nsch T~W and Bloch I 2002 {\em nature\/} {\bf 415} 39--44

\bibitem{tang2015ground}
Tang G, Eggert S and Pelster A 2015 {\em New Journal of Physics\/} {\bf 17} 123016

\bibitem{zhang2017ground}
Zhang W, Greschner S, Fan E, Scott T~C and Zhang Y 2017 {\em Physical Review A\/} {\bf 95} 053614

\bibitem{vishveshwara2008josephson}
Vishveshwara S and Lannert C 2008 {\em Physical Review A—Atomic, Molecular, and Optical Physics\/} {\bf 78} 053620

\bibitem{josephson1962superconductive}
Josephson B 1962 {\em Physics Letters\/} {\bf 1} 251--253 received 8 June 1962

\bibitem{wiki_josephson_effect}
contributors W 2024 Josephson effect --- wikipedia{,} the free encyclopedia accessed: 2024-11-22 \urlprefix\url{https://en.wikipedia.org/wiki/Josephson effect}

\end{thebibliography}

\end{document}